# ON THE AUTOMATED PLANNING AND DESIGN OF SMATV SYSTEMS

Radu ARSINTE
*Technical University of Cluj-Napoca, Communications Department*
*26-28 Baritiu Str., radu.arsinte@com.utcluj.ro*

**Abstract:** The paper presents some theoretical and practical considerations regarding the TV information distribution in local (small and medium) networks, using different technologies and architectures. The SMATV concept is chosen to be presented extensively. The most important design formulae are presented with a software package supporting the network planner to design and optimize the network. A case study is realized, using standard components in SMATV, for a 5 floor building. The study proved that it is possible to design and optimize the entire network, without realizing first a costly experimental setup. It is also possible to run different architectures, optimizing also the costs of the final solution of network.

## I. SYSTEMS FOR COLLECTIVE USE OF SATELLITE SIGNALS [1][2]

Satellite TV, as well as air and cable starts to become increasingly important for the residents of crowded cities. Satellite receiving antennas are already found on the walls of urban high-rise buildings and on the roofs of houses and village houses. This is reminiscent of the period of the 60s, when every owner had his own TV antenna. Acquiring the necessary equipment isn't all. It is necessary to take into account several additional factors: your apartment should not be on the ground floor, its windows should not go in the "yard-well", and at least one window should go to the south or in the worst case, to the west or east. Complying with all of these recommendations is not always possible for many reasons (financial, aesthetic, technical), so it is more often a rational creation of a system of collective reception.

Among the most popular ways of delivery SAT channels to subscribers are:

- Traditional method with a full signal demodulation to the audio/video (A/V) followed by conventional analog modulation using amplitude modulation (AM-TV).
- SMATV systems using multi-switchers.
- The combined method comprising simultaneously broadcast on the same cable as in the range of CATV (47-862 MHz) ATV and DVB signals (channel bandwidth BW = 8 MHz) and a direct SAT distribution channels in the intermediate frequency (IF) range of 950-2150 MHz (BW = 27/36 MHz).
- Direct SAT distribution channels through CATV in a frequency range of the 47 - 862 MHz converted back to an IF frequency range of 950-2150 MHz.
- Using trans-modulation DVB-S/S2 to DVB-C/T.
- Direct distribution in HFC with/without frequency conversion.
- Reformatting in MPEG streams using IP format and distribution on a digital IP networks.

SMATV (Satellite Master Antenna Television) [3] is a satellite TV system for collective use, providing an autonomous solution for different channels with multiple antennas (the actual number is determined by the frequency range and conditions of the area). When SMATV is used, the costs to receive satellite TV programs will be reduced by almost half. Figure 1 shows that the subscriber's personal network should contain only a satellite receiver, the cost for installation and acquisition antennas are divided for all subscribers. Subscribers can be, depending on the equipment, up to 50 people (200 in other references), that is enough in most cases. One system can handle, for example, one or two entrances of a multiple module building. Furthermore, it may be extend the signal broadcast of local TV broadcasting. The subscriber is able to self-and independent choice of programs broadcast satellite or satellites (as satellite dishes may be two, or one antenna may take the signals of two satellites at a time). There are strict requirements that apply not only to the antenna-feeder and amplifying equipment, television receiver systems, but the elements distribution networks, providing the branching of TV signal and division of its energy. The main difference of collective and individual receiving of television programs is associated with quality requirements of the received image.

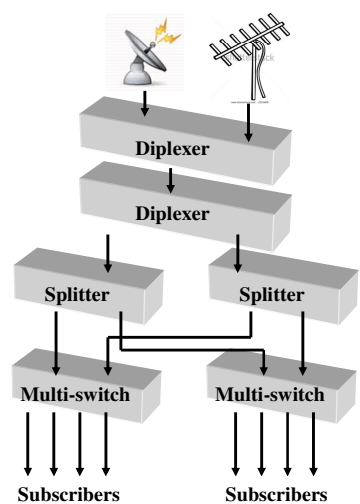

*Figure 1. Conceptual SMATV distribution network*

Thus, the energy quality factor for receiving stations in collective reception systems usually must be at least 14dB/K and S/N ratio in TV channel of the cable not less than 57 dB. This forces to use in the collective receiving complex an antenna with a large diameter. Parameters of receiving systems and the prospects for their improvement are largely dependent on the quality and reliability used couplers and dividers. The main objective of user taps a power distribution RF signals between the taps. Subscriber taps should not interfere with television receivers, producing interferences by their local oscillators, and if the frequency of the signal LO or their harmonics are not in the frequency bands of channels from the distribution network, it is possible a separation between the taps of 20 dB, but for multi-channel receiver networks, it should be less than 40 dB. The own loss of couplers is a serious matter, since a large number of distribution networks use them in series. Depending on the composition of receiving equipment, complex communities can receive not only satellite, but also terrestrial TV programs and FM radio (5 - 2150 MHz). To create such a distribution system needs a large amount and a variety of devices: switches, diplexers, subscriber splitters outlets, amplifiers, attenuators, multi-switches et al. (see. Figure 2). The use of such devices has allowed developing a number of simple and reliable circuits networks capacity is up to 50 analogue TV channels.

The introduction of new standards of DVB (second generation), has revitalized the SMATV research, to accommodate the networks with the technical requirements of the next DVB generation. ([4], [5])

**II. DESIGN PRINCIPLES OF SMATV SYSTEMS**

If necessary, the satellite programs are captured simultaneously in two polarizations and/or two sub-bands, as well as from several satellites, and special devices (multi-switches) are used for routing the inputs. Such a switch in the simplest case, consist of a matrix switch with N inputs and M outputs for the subscribers. The number of SAT inputs (950-2150 MHz) of multi-switches usually varies from 2 to 16. As a rule, a part of the multi-switch is a terrestrial TV input providing an additional active or passive input for terrestrial TV (47 - 862 MHz), which allows you to broadcast all the signals (TV and SAT) on a single cable. Each subscriber can independently select the desired SAT entry by applying a control voltage from your tuner. Typically, the control is performed due to changes in the value of the supply voltage of the low noise converter (LNC) 13/18V, the frequency of the supply voltage 0/22kHz or using the digital control protocol DiSEqC.

All multi-switches can be divided into 3 classes:
1. Stand alone (autonomous) - an independent use of individual subscribers. Sometimes they are used successfully and as final multi-switches.
2. Cascadable - used in collective SMATV, have not only inputs and drop cables, but also outputs (inputs and outputs are always the same). It may contain reap the plug-in power supply or feed themselves from the preceding or post-blowing in cable multi-switches.
3. Terminal (final cascade) - do not have outputs. Most often

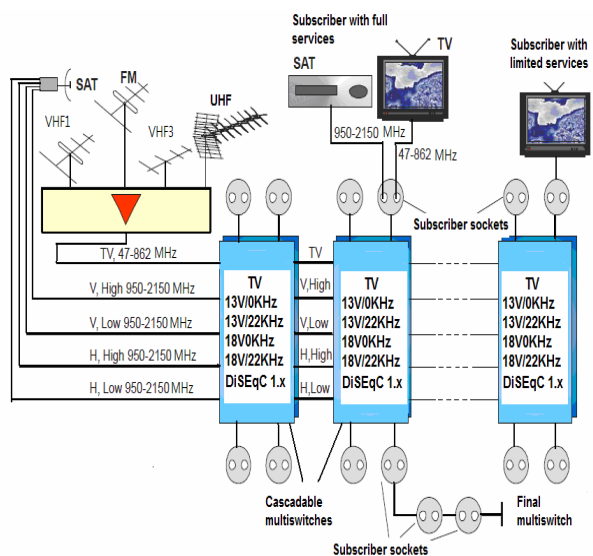

*Figure 2. Technical diagram of a SMATV system [6]*

they do not contain a supply source.

The principle of multi-switches inclusion in a complex system with amplification in TV and SAT ranges (for connection of up to 200 subscribers) is shown in Figure 2. Outlets for such a hybrid network are equipped with at least two exits - to connect SAT receiver (F connector) and TV (IEC type). But can be used many other types of output receptacles. The design of such complex systems is ruled by many design rules and formulae. In most cases there are simplified procedures producing rapid results with a good approximation. Here is a short list of practical rules for a SMATV network.

*Rule 1*

If the signal provided by the equipment manufacturer within the power criteria, for example, in dBm, then the conversion to the usual value of the voltage (dBmV or dBµV) per channel should be carried out according to the formula:

$$U_{out} = P_{dBmW} = 108 - 10\lg(N) \qquad (1)$$

where N - number of SAT channels. For example, if Pmax = 0 dBm, the maximum allowable signal level for the translation of 2 channels will be 105.8 dBµV, and for 30 channels would be equal to 94 dBµV.

*Rule 2*

If the project costs allows, it is desirable to broadcast in SMATV only digital packages that you need. It is no secret that interesting programs are duplicated in different packages on different satellites. Moreover, very often there are the cases when all over the full range of 950-2150 MHz we only need 1 or 2 packets.

It is necessary to translate all the packages at once. For the selection of channels exist SAT headends (GS) of the class IF-IF. These headends are intended to select only the transponders in which you are interested in and put them in the network without any conversion, i.e. preserving the original modulation format.

*Rule 3*

Do not compromise the system by the noise accumulation (i.e., reduce C/N) in the translation of SAT channels, by the reduced dynamic range of the amplifier, typically 2-3 orders

___

of magnitude above the translated C/N. Indeed, realized (final) value of C/N at the output of the chain consisting of N identical amplifiers is determined by the known formula:

$$C/N = -10\lg\left(10^{-(C/N)_{out}/10} + 10^{-(U_{out}-K-F-10)/10} - 10\lg(N)\right) \quad (2)$$

where: $C/N_{in}$ - input ratio of C/N (e.g., to the output of SAT headend or LNB output);
$V_{out}$ - the signal levels at the amplifier output;
K - gain of amplifier;
F - noise factor of the amplifier.
For example, when $C/N_{in}$ = 16 dB (a rather high value) and there are 10 cascaded amplifiers (an extended tree) with $V_{out}$ = 80 dB V (output levels are lowered on purpose in order to preserve IMD) at K = 36dB, and F = 8 dB, the output C/N is 15.6 dB, i.e. decreases by only 0.4 dB. In this case, the assumptions for the example adopted are establishing strict conditions.

## III. SOFTWARE FOR ASSISTED DESIGN OF LOW COMPLEXITY SMATV NETWORKS

As seen in the previous paragraph, the design of SMATV distribution networks requires the consideration of many rules and nonlinear formulae. To simplify the process and handle a large complexity of devices (passive and active) in the process of selection and design, most cable devices manufacturing companies are developing special software packages. This is particularly important for engineers and technicians will less expertise in the theoretical field of RF circuit design, allowing fast and standardized designs, ready to be implemented will minimal costs.

An example is the SatNet [7] program. This software is intended to assist the planner in the designing of SAT IF distribution networks based on multi-switches.

The design assisted by SatNet has basically the following steps:
- Establishing the main requirements of the network: number of floors, number of subscribers per floor, number of inputs for each subscriber (SAT and TV);
- Editing the network using the integrated editor and the attached database with standard components;
- Setup the initial conditions of the design;
- Run the simulation to identify the weak spots of the network (outputs with signals outside the limits);
- Make de necessary corrections (gain modifications, cable lengths, eventually component changes) to comply with design conditions;
- Finalization of the design.

## IV. DESIGN EXAMPLE

Figure 3 contains an example of SMATV design for a condominium. The design assumes a 5 floors building, with 4 apartments on each floor, and 2 SAT links and 1 TV connection for each one. We must mention that TV connections can be multiplied in each apartment, SAT connections not.

The design is realized using the modules presented in [8] that are contained and introduced as basic models in the software package.
The software has an advanced graphical editor, making possible to easily build the entire network.
The network contains basically the following components.
*MV5xx - Remotely powered multiswitches*
- for large installations of SAT IF distribution systems
- powered in line through H lines
- compatibility of all components from cable distribution system
- subscriber line length up to 80 meters
- gain regulator with four discrete positions for each SAT IF line and separate 16 positions discrete gain regulator for terrestrial TV
- optimized for operation with terrestrial digital/analog signals
- active terrestrial TV path is powered from central power supply allows to receive terrestrial TV programs without switching on SAT TV receiver
- LED indication of 18 V line powering
- supply power for SAT LNBs

*MR512- Radial multiswitches*
- intended for star distribution system of 4 SAT IF polarities and terrestrial TV signal up to 16 users
- discrete gain regulator with 16 positions for terrestrial TV
- built-in power supply for remote DC feeding
- possibility to apply DC power to the preamplifier through terrestrial TV input

*SD5xx - Taps and splitter*

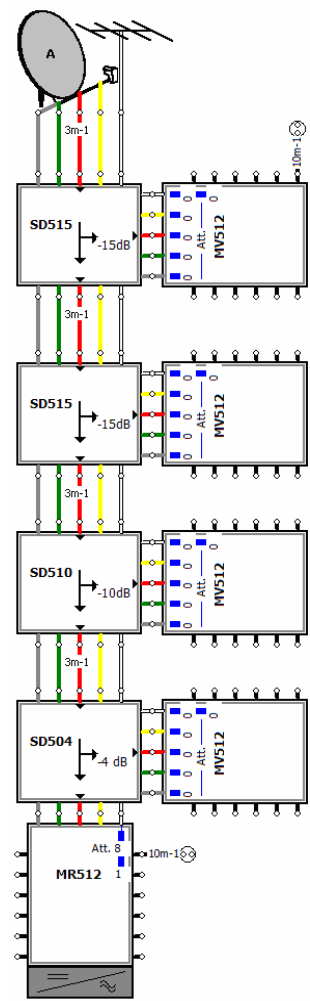

*Figure 3.The proposed design example*

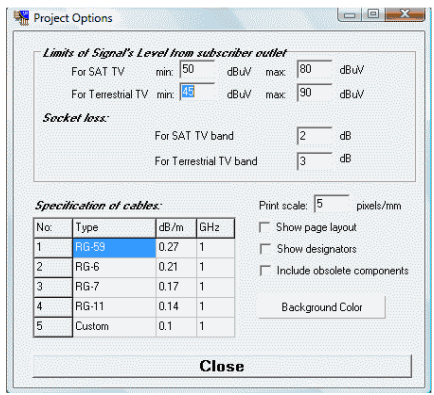

*Figure 4. Setting the initial conditions of the design*

- 2 way splitter and one way taps of 4 SAT+1 terrestrial signals
- Losses of 4-15 dB
- DC pass through SAT and terrestrial TV trunk lines;
- switchable DC pass to tap H outputs

Once introduced the network with the editor, the initial conditions of the design must be set. In the window (Figure 4), the constraints of the design process are established, and the parameters of the cables used in the design. The levels of the signals for the subscriber outlets are established in the technical standards [3], but must be confronted with actual specifications of SAT receivers and TV sets used by subscribers.

Most satellite SAT-IF systems for the distribution of signals are cascade-type or have an extended branch of multi-cascaded switches. Beside the well-known benefits of these networks, they have also disadvantages. For example, it is difficult to balance perfectly the signal levels at subscribers and at the same time to avoid overloading of active network components (amplifiers, multi-switches). The first way to overcome this problem is the gain adjustment (loss/gain), mounted on many active components. At this moment the simulation can be launched, to verify the degree of compatibility of the design with the established goals.

The first run (with default parameters) shows an overloading of most terrestrial outputs of the 1-5$^{th}$ floors. Adjusting the levels of the terrestrial TV input signal makes possible to arrange most outputs to an acceptable level. Some outputs remain in a defective state, like the characteristic presented in figure 5, where the output is trespassing the limits for some frequencies. Modifying the signal input of terrestrial TV it is possible to correct the characteristics obtaining the values from table 1.

Table 1.

| Input level terr. TV [dB] | Number of outputs within limits | Number of outputs outside limits |
|---|---|---|
| 50 | 0 | All |
| 60 | 0 | All |
| 70 | 24 | 36 |
| 80 | 57 | 3 |
| 90 | 4 | 56 |

Analyzing the results from table 1 it is possible to observe that the optimal level is obtained for signals around 75-85 dB. This condition can be achieved by adding an attenuator (or an adjustable gain amplifier) to the terrestrial TV input. The software offers an optimization tool, making possible to adjust the gain of the amplifiers in order to raise the number

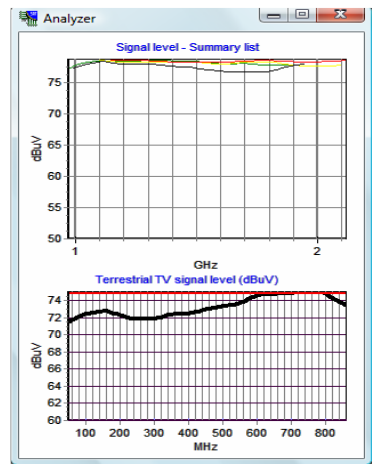

*Figure 5. Signal characteristics of an output (Up- SAT, Down- Terr. TV)*

of outputs within limits.

**V. CONCLUSIONS**

We have presented in the paper some theoretical considerations regarding the TV information distribution in local (small) cable networks, using SMATV concepts.

The most important design formulae are presented with software supporting the network planner to design and optimize the network. A case study is realized, using standard components in SMATV, for a 5 floor building.

The study proved that it is possible to design and optimize the entire network, without realizing first a costly experimental setup. The adjustments possible on each component makes possible to accommodate later the design to minor modifications required during and after installation. The most important element is the fact that the design can be realized without analyzing the actual internal structure of the SMATV components, i.e. at the conceptual level.

We are intending to extend the study to more complex structures and designs in TV for systems using alternative methods of TV distribution, including or not, full SMATV technologies.